\documentstyle[12pt]{article}
\begin{document}
\begin{titlepage}
\rightline{SNUTP/95-113}
\def\today{\ifcase\month\or
        January\or February\or March\or April\or May\or June\or
        July\or August\or September\or October\or November\or December\fi,
  \number\year}
\rightline{hep-th/9510242}
\rightline{November, 1995}
\vskip 1cm
\centerline{\Large \bf Yangian Symmetries in the $SU(N)_1$ WZW Model}
\centerline{\Large \bf and the Calogero-Sutherland Model}
\vskip 2cm
\centerline{{\large {\sc Changhyun Ahn}} and
{\large {\sc Soonkeon Nam}} \footnote{
nam@nms.kyunghee.ac.kr}}
\vskip 1cm
\centerline {{\it Department of Physics, Kyung Hee University}}
\centerline {{\it and Research Institute for Basic Sciences}}
\centerline {{\it Seoul 130-701, Korea}}
\vskip 1cm
\centerline{\sc Abstract}
\vskip 0.2in
We study the $SU(N)$, level $1$ Wess-Zumino-Witten model, with
affine primary fields as spinon fields of fundamental representation.
By evaluating the action of the Yangian generators $Q_{0}^{a},
Q_{1}^{a}$ and the Hamiltonian $H_2$ on two spinon states we get a new
connection between this conformal field theory and the Calogero-Sutherland
model with $SU(N)$ spin. This connection clearly confirms the need for the
$W_3$ generator in $H_2$ and an additional term in the $Q^{a}_{1}$. We also
evaluate some energy spectra of $H_2$, by acting it on multi-spinon states.
\end{titlepage}
\newpage

\def\beq{\begin{equation}}
\def\eeq{\end{equation}}
\def\bea{\begin{eqnarray}}
\def\eea{\end{eqnarray}}
\renewcommand{\arraystretch}{1.5}
\def\ba{\begin{array}}
\def\ea{\end{array}}
\def\bce{\begin{center}}
\def\ece{\end{center}}
\def\nn{\noindent}
\def\nonu{\nonumber}
\def\pbx{\partial_x}


\def\ptl{\partial}
\def\al{\alpha}
\def\be{\beta}
\def\ga{\gamma}
\def\Ga{\Gamma}
\def\de{\delta} \def\De{\Delta}
\def\ep{\epsilon}
\def\vep{\varepsilon}
\def\ze{\zeta}
\def\et{\eta}
\def\th{\theta} \def\Th{\Theta}
\def\vth{\vartheta}
\def\io{\iota}
\def\ka{\kappa}
\def\la{\lambda}
\def\La{\Lambda}
\def\rh{\rho}
\def\si{\sigma} \def\Si{\Sigma}
\def\ta{\tau}
\def\up{\upsilon}
\def\Up{\Upsilon}
\def\ph{\phi}
\def\Ph{\Phi}
\def\vph{\varphi}
\def\ch{\chi}
\def\ps{\psi}
\def\Ps{\Psi}
\def\om{\omega}
\def\Om{\Omega}

\def\lbr{\left(}
\def\rbr{\right)}
\def\half{\frac{1}{2}}
\def\CVO#1#2#3{\!\left( \matrix{ #1 \cr #2 \ #3 \cr} \right)\!}

\def\vol#1{{\bf #1}}
\def\nupha#1{Nucl. Phys. \vol{#1} }
\def\phlta#1{Phys. Lett. \vol{#1} }
\def\phyrv#1{Phys. Rev. \vol{#1} }
\def\PRL#1{Phys. Rev. Lett \vol{#1} }
\def\prs#1{Proc. Roc. Soc. \vol{#1} }
\def\PTP#1{Prog. Theo. Phys. \vol{#1} }
\def\SJNP#1{Sov. J. Nucl. Phys. \vol{#1} }
\def\TMP#1{Theor. Math. Phys. \vol{#1} }
\def\ANNPHY#1{Annals of Phys. \vol{#1} }
\def\PNAS#1{Proc. Natl. Acad. Sci. USA \vol{#1} }

In recent years, a number of Conformal Field Theories (CFT's) have been
reconstructed in massless quasi-particle bases of CFT's,
among many equivalent
bases of CFT's. This is better suited for the description of particles
in the massive perturbations of CFT's than in the conventional Verma-module
basis.
One of very interesting examples which can be studied with
the quasi-particle description is the low energy limit of the
Calogero-Sutherland (CS) model with
$SU(2)$ spin\cite{HHTBP}.
This new formulation does not resort to affine and Virasoro symmetries, but
to Yangian symmetry. The Hilbert space may be regarded as spinon Fock space
constructed using the generalized commutation relations\cite{BLS1}.
It is now known that Yangian generators can be written in terms of
$SU(2)$, level $1$ affine currents.
Furthermore the action of these generators on the
multi-spinon states provides a close relation with the low energy limit
of the CS model with $SU(2)$ spin\cite{BPS}.
The generalization to higher level
$SU(2)$ Wess-Zumino-Witten (WZW)
model has also been constructed in Ref.\cite{BLS2}. However, for higher level
cases, it is impossible to represent the Yangian generators in terms of
affine current.

Regarding the $SU(N) (N>2) $, level 1 generalization, the additional structures
due to
the conformal dimension $2$ and $3$ fields are inevitable.
It has been found in Ref.\cite{Sch} that the Yangian
generator $ Q_{1}^{a} $ and the Hamiltonian $ H_{2} $ need extra terms
in terms of $W^a=\frac{1}{2}d^{abc}(J^b J^c)$ and the third order
Casimir operator $W_3= \frac{1}{6} d^{abc} (J^a (J^b J^c))$
(notations for these quantities will be explained later).
However it has not been demonstrated that the action of the
Yangian generator and the Hamiltonian are {\it indeed} those of the
CS model with $SU(N)$ spin.
This is an interesting problem in light of the work in Ref.\cite{BGHP},
where the $SU(N)$ generalization of the Hamiltonian of the
dynamical CS model and its Yangian structure was discussed.
Another interesting problem is the notion of spinons for $SU(N)$, $N>2$.
In the study of Heisenberg spin chains, Reshetikhin\cite{Reshet}
observed that there are in general $N-1$ types of spinons for $SU(N)$
case. So we expect similar thing in our case. To fully describe the
energy spectrum of the model, one has to introduce $N-1$ species of
spinons.

In this letter, we consider the $SU(N)_{1}$ WZW CFT and construct the Yangian
generators and the Hamiltonian $H_2$
in terms of the affine currents of the theory.
To determine the coefficients of the extra terms $W^a$ and $W_3$
in $Q^a_1$ and $H_2$ one can check the Yangian algebra $Y(sl_N)$ and the
commutativity of $ [ Q_1^a, H_{2} ] =0 $, which are
very complicated \cite{Sch}.
Rather, we apply them to
multi-spinon states of $\overline{N}$-representation,
and compare the result with the Hamiltonian and the Yangian generators
in the dynamical model\cite{BGHP}.
This uniquely fixes the coefficients and gives further physical motivation
for the $W$-structures.
In general there can be spinons of representation other than the fundamental
representation, but these are not considered
in this letter.

The Yangian generator $Q_1^a$ and the Hamiltonian $H_2$
consist of two parts, one of which contains
the derivative terms, $\sum_{i} D_i$ and $\sum_{i}D^2_i$,
and the other part contains the interaction term,
$\sum_{i \neq j} t_i t_j$, where $D_i = w_i\partial_{w_i}$, $w_i$ being
the position of the $i$-th particle in the CS model, and $t_i$ the $SU(N)$
generator acting on the $i$-th particle.
We observe that without the contributions from $W^a$ and $W_3$,
the derivative part is smaller than that is required from integrability.
That is, in CS model the relative coefficient between the derivative term and
interaction term is fixed {\it uniquely} by integrability\cite{Poly}.
On the other hand, $W^a$ and $W_3$ contribute only to
the derivative part up to irrelevant constant terms,
and we can therefore determine the coefficients in front of
them from the integrability.

The $SU(N)_1$ WZW model has the affine Kac-Moody current $ J^a(z) $ which
satisfies the operator product expansion (OPE),
\beq
J^{a} (z) J^{b} (w) = \frac{\de^{ab}}{(z-w)^2} + \frac{1}{(z-w)} f^{abc}
J^{c} (w) + \cdots,
\eeq
where $a =1,2, \cdots ,N^2-1$, and $ f^{abc}$ are the structure constants of
$SU(N)$ with the normalization $f^{abc} \; f^{bcd} = -2N \de^{ad}$. We do not
write down explicitly the regular parts of OPE.
The chiral vertex operators of $\overline{N}$-representation of
this model are identical as
the spinon fields $\ph^{\al}(z)$, $\al =1,2, \cdots, N, $ which transform
according to
\beq
J^{a} (z) \ph^{\al} (w) = \frac{ {(t^a)^{\al}}_{\be} \; \ph^{\be}(w)}
{(z-w)} + \cdots,
\label{eq:jp}
\eeq
where ${{(t^a)}^{\al}}_{\be}$ are the $\al \be$ elements of the
$SU(N)$ matrices in the fundamental
representation, $[t^{a}, t^{b}] = f^{abc} t^{c}$.

The first two generators of $Y(sl_2)$ Yangian can be written
in terms of the affine Kac-Moody current.
This can be motivated
from a limiting procedure which takes the number of lattice sites of a
spin system to infinity\cite{Haldane}. In case of $Y(sl_N) (N>2)$
one has to be more careful since new local operators can arise from this
limiting procedure\cite{Sch}. There is of course
no correction for $Q^a_0$.
Thus we write down the Yangian generators in terms of affine currents
and also a contribution from $W^a$ as follows:
\bea
Q^a_0 &=& J^a_0, \nonu \\
Q^a_1 &=& Q^a_{1,{\rm unc}} + \mu \; Q^a_{1,{\rm corr}}  \nonu \\
&=& \frac{1}{2} f^{abc} \sum_{m > 0} \lbr J_{-m}^{b} J_{m}^{c} \rbr
+ \mu \; d^{abc} ( J^{b} J^{c} )_{0},
\eea
where $J^a_m, m \in Z $ is the Laurent mode of $J^a(z)= \sum_{m}
J^a_m z^{-m-1}$ and $ d^{abc} $ is the completely symmetric tensor,
$ \{ t^a, t^b \} = d^{abc} t^c + \frac{2}{N}\de^{ab} $ with the normalization
$ d^{abc} d^{bcd}=\frac{2(N^2-4)}{N} \de^{ad}$.
The brackets mean normal ordering between
fields\cite{BS}. In principle, we may think of $ f^{abc} (J^b J^c)_0$ as the
other possible candidate for correction term. But this is proportional to
the zero mode of $\partial J^a$ which does not give rise to the derivative
part of $Q_1^a$.
Similarly, the Hamiltonian $H_2$ can be written as follows:
\bea
H_2&=& H_{2,{\rm unc}} + \nu \; H_{2,{\rm corr}}\nonu \\
&=& \sum_{m > 0} ( m J_{-m}^{a} J_{m}^{a} )+\nu \;
d^{abc} (J^{a} ( J^{b} J^{c} ))_{0}.
\eea
In the above `unc' and `corr' stand for uncorrected and
corrected parts, where
we mean by the uncorrected term which is the same form as in $Y(sl_2)$.
The parameters, $\mu, \nu$ will be determined later.

First of all, we consider the action of $H_{2}$ on the two spinon states
given by the following contour integrals:
\bea
&&H_{2} \; \ph^{\al} (w_{2}) \ph^{\be} (w_{1}) | 0 \rangle  =
\left[ \oint_{|z_{1}| > |z_{2}|} \frac{dz_{1}}{2 \pi i}
\oint_{C_{i}} \frac{dz_{2}}{2 \pi i} \frac{z_{1} z_{2}}{(z_{1}-z_{2})^2}
J^{a}(z_{1}) J^{a}(z_{2}) \right.   \nonu \\
&&\hskip 2cm \left.+ \nu \; d^{abc} \oint_{C_{i}} \frac{dz}{2 \pi i} z^2
( J^{a} ( J^{b} J^{c} ) ) (z) \right]
\ph^{\al} (w_{2}) \ph^{\be} (w_{1}) | 0 \rangle  \nonu \\
&&\hskip 3.5cm =  \left[ \frac{N^2-1}{2 \De ( 2 \De +1) N} \left( 1+ \nu \;
\frac{6(N^2-4)}{N}
\right) \sum_{i} \left( (D_{i})^2 +2 \De D_{i} \right) \right. \nonu \\
&&\hskip 2cm \left. - \sum_{i \neq j} \th_{ij}
\th_{ji} t_{i}^{a}  t_{j}^{a} + 2 \nu \; C_3(\overline{N}) \right]
\ph^{\al} (w_{2}) \ph^{\be} (w_{1}) | 0 \rangle,
\label{eq:h2}
\eea
where
the contours $C_{i}$'s  encircle the points $w_{1}, w_{2}$ and
$D_{i} =w_{i} \partial_{w_{i}}, \th_{ij}=\frac{w_{i}}{w_{i}-w_{j}}$ and
$C_3(\overline{N})=\frac{(N^2-1)(N^2-4)}{N^2}$.
We used the following fact;
\beq
{(t^{a} \; t^{b})^{\al}}_{\be} = {\left( \frac{\de^{ab}}{N} +\half f^{abc}
t^{c} +\half d^{abc} t^{c} \right)^{\al}}_{\be},
\label{eq:tt}
\eeq
and many identities among $f^{abc}$ and $d^{abc}$. ( See Ref.\cite{BS}
for conventions. )
We can raise and lower spinor indices with antisymmetric symbols
$\ep_{\al \be}$ and $\ep^{\al \be}=-\ep_{\al \be}$.

The correction term vanishes automatically for $N=2$  and the rest
is in agreement with the one in Refs.\cite{BPS,BLS2}.
In this calculation, one should know the normal ordering of $
( J^{a} \ph^{\al} ) (z) $, which is the next term of the right hand side
of Eq.(\ref{eq:jp}).
It
can be written as
\beq
( J^{a} \ph^{\al} ) (z) = \frac{1}{ 2 \De} {( t^{a} ) ^{\al}}_{\be} \partial
\ph^{\be} (z),
\eeq
where $\De = \frac{N^2-1}{2N(N+1)}$ is the conformal dimension of the
primary field $\ph^{\al}(z)$\cite{KZ}. One can evaluate explicitly the OPE
between stress-energy tensor in Sugawara form
$ T= \frac{1}{2(1+N)}
(J^{a} J^{a} ) (z) $ (with the central charge $c=N-1$),
and spinon field $\ph^{\al} (w)$,
according to Eq.(\ref{eq:jp}) and recognize that $\De$ is the coefficient of
$\frac{1}{(z-w)^2} \ph^{\al} (w)$.
The mode of $T(z)$, i.e. $ L_{n}, $ is given by
\beq
L_{n}=\frac{1}{2(1+N)} \sum_{m=-\infty}^{\infty} ( J_{m}^a J_{n-m}^a ).
\eeq
By acting the spinon field $\phi^{\al}$ on both sides, we have
the following relation:
\beq
L_{-2} \phi^{\al} = \frac{1}{4 \De} \partial^2 \phi^{\al}+
\frac{N(4 \De-1)}{2(N^2-1)}
t^a J^{a}_{-2} \phi^{\al}.
\eeq
Then the normal ordering of $(\partial J^{a} \phi )(z)=J^{a}_{-2} \phi(z)$
leads to
\beq
(\partial J^{a} \phi^{\al})(z) = \frac{1}{2 \De(2 \De+1)} t^{a} \partial^2
\phi^{\al}(z)+
\frac{2(N^2-1)}{N(2N-1)} \; \Psi^{a,\al}(z),
\label{eq:null}
\eeq
where
\beq
\Psi^{a, \al}(z) \equiv
(\partial J^{a} \phi^{\al})(z) -\frac{N^2}{2(N^2-1)} t^a
\partial^2 \phi^{\al}(z)+
\frac{N}{N^2-1} t^{a} \left( T \phi^{\al} \right)(z),
\eeq
which is a primary field of conformal dimension $ 2+\De=
\frac{5N-1}{2N} $.
The second term of Eq.(\ref{eq:null}) vanishes when
we calculate $ t^a (\partial J^{a} \phi^{\al})(z)$.
Finally we get the closed form of Eq.(\ref{eq:h2}) by exploiting the above
arguments. We would like to stress that the contribution from correction term
does not give rise to the interaction term,
$t_i^a t_j^a$, but the derivative
terms, $D_i$ and $D_i^2$.
This is due to the fact that this corrected term is purely
the {\it zero mode} of a conformal field in the theory. In other words, it acts
on only one spinon field, not on two.

The Hamiltonian of the dynamical model\cite{Poly} is given by
\bea
H_{2}^{dyn} & = & \sum_{i} (D_{i})^2 +\sum_{ i \neq j}
\la ( P_{ij} + \la )
\th_{ij} \th_{ji} \nonu \\
& =  & \sum_{i} (D_{i})^2+\sum_{i \neq j} \la ( t^{a}_{i} t^{a}_{j} +
\frac{1}{N} + \la ) \th_{ij} \th_{ji},
\label{eq:dyn}
\eea
where we used that the permutation operator $P_{ij}$  for $SU(N)$
can be written as
\beq
P_{ij} = t^{a}_{i} t^{a}_{j} +\frac{1}{N}.
\eeq
We can fix the parameter $ \nu$ by identifying the Hamiltonian in
Eq.(\ref{eq:h2})
with that in Eq.(\ref{eq:dyn}) up to an overall factor.
Then, we must first have,
\beq
\lambda = -\frac{1}{N},
\eeq
and also
\beq
\frac{N^2-1}{2 \De ( 2 \De +1) N} \left( 1+ \nu \;
\frac{6(N^2-4)}{N}
\right) =N.
\label{eq:con}
\eeq
So we get
\beq
\nu= \frac{N}{6(N+1)(N+2)}.
\eeq
This coefficient is exactly the same as the one in Ref.\cite{Sch}.
It is important to notice that the reason for the extra $\nu$ term is
evident in Eq.(\ref{eq:con}).
Putting these together, the Hamiltonian $H_{2}$ acts on two spinon
states by a linear combination of
$H_{1}^{dyn}=\sum_{i} D_{i}$ and $H_{2}^{dyn}$ as follows:
\beq
H_{2}=\frac{(N-1)(N-2)}{3N} + N \left( H_{2}^{dyn} + 2 \De H_{1}^{dyn}
\right),
\eeq
with
\beq
\la = -\frac{1}{N}.
\eeq

The action of $Q^a_1$ on two spinon states is given by
the following contour integrals:
\bea
&&Q_{1}^a \; \ph^{\al} (w_{2}) \ph^{\be} (w_{1}) | 0 \rangle =
\left[ \half \oint_{|z_{1}| > |z_{2}|} \frac{dz_{1}}{2 \pi i}
\oint_{C_{i}} \frac{dz_{2}}{2 \pi i} \frac{z_{2}}{z_{1}-z_{2}}
f^{abc} J^{b}(z_{1}) J^{c}(z_{2}) \right. \nonu \\
&& \hskip 2cm \left. +
\mu \;  d^{abc} \oint_{C_{i}} \frac{dz}{2 \pi i} z \left(
J^{b} J^{c} \right) (z) \right] \ph^{\al} (w_{2}) \ph^{\be}
(w_{1}) | 0 \rangle,
\label{eq:yan}
\eea
where the contours $C_{i}$'s encircle the points $w_{1}, w_{2}$.
We arrive at the following expressions after doing the integrals:
\bea
&&Q_{1}^a \; \ph^{\al} (w_{2}) \ph^{\be} (w_{1}) | 0 \rangle =
 \left[ \left( -\frac{N}{4\De} +\mu \; \frac{N^2-4}{\De N}
\right) \sum_{i} D_{i} t_{i}^{a} \right. \nonu \\
&&\hskip 2cm \left. + \mu \;
\frac{N^2-4}{N} Q_{0}^{a} +
\half f^{abc} \sum_{i \neq j} \th_{ij} t_{i}^{b} t_{j}^{c} \right]
\ph^{\al} (w_{2}) \ph^{\be} (w_{1}) | 0 \rangle.
\label{eq:qpf}
\eea
Again the correction terms vanish for $N=2$.
The contribution due to the presence of second term in Eq.(\ref{eq:yan})
does not give rise to the last term of Eq.(\ref{eq:qpf}). It is easy to
check that the Yangians given by the differential operators satisfy
Yangian algebra.
Comparing with the Yangian generator of dynamical
model given in Ref.\cite{BHW}, we can determine the
parameter $\mu$ as follows using the relation $\la=-\frac{1}{N}$ obtained in
Eq.(\ref{eq:con})
\beq
 -\frac{N}{4\De} +\mu \; \frac{N^2-4}{\De N} = \frac{1}{\la} ( =- N ).
\eeq
That is, we must have
\beq
\mu = - \frac{N}{4(N+2)}.
\eeq
This gives an independent confirmation of the Ansatz given
in Ref.\cite{Sch}, where the coefficient was fixed by the Yangian algebra
$(Y1)-(Y4)$ and the commutativity of $[ Q^a_1, H_2 ] = 0$.
Here checking the Yangian algebra in terms of differential operators is
easier than in terms of affine currents.

The final form of $Q_1^a$ on two spinon states is given by
\bea
&&Q_{1}^a \; \ph^{\al} (w_{2}) \ph^{\be} (w_{1}) | 0 \rangle =
 \nonu \\
&&\hskip 1cm \left[ -N \sum_{i} D_{i} t_{i}^{a}
+\half f^{abc} \sum_{i \neq j} \th_{ij} t_{i}^{b} t_{j}^{c} -
\frac{N-2}{4} Q_{0}^{a} \right]
\ph^{\al} (w_{2}) \ph^{\be} (w_{1}) | 0 \rangle.
\label{eq:qpf1}
\eea
Just as in the $H_2$ case, the term of $ -\frac{N-2}{4} Q_0^a$ can be
absorbed in $Q_1^a$ by redefining it because it induces only a
spectral parameter shift.

Having found the differential forms of the Hamiltonian and the Yangian
generators, we now look for the values of low lying energy spectrum of
$H_2$.
It was found in Ref.\cite{HHTBP} that the eigenfunctions of the models
are grouped into multiplets characterized by sets of distinct positive
integers $\{m_i\}$, called motifs, which can be transformed into
sequences of 0's and 1's.
The rules on the sequences are that at most $N-1$ consecutive 1's
can occur, and that the asymptotic behavior of the sequence
falls into $N$ distinct cases, each of which corresponds to the $N$
primary sectors including the identity, of the $SU(N)$ level 1 WZW model.
Some of low lying energy levels of $SU(3)$ were found in Ref.\cite{Sch}.
We will recover these energy levels by considering the actions of
$H_2$ and $H_1$ on states with various modes of spinon fields
of fundamental representation acted on
the vacuum $|0\rangle$.

We can define a spinon field of representation $\lambda$
corresponding to Young diagram of $\left[1\right]^r (r=1,2, \cdots N-1)$
as chiral vertex operators transforming in the irreducible
$sl_N$ representation of $\lambda$, i.e.
$\Ph\ : \ L_{\si}\rightarrow \L_\rh$.
The mode expansion of the spinon is given by following:
\beq
\Phi \CVO{\la}{\rh}{\si} (z)= \sum_{n \in Z}\Phi
\CVO{\la}{\rh}{\si} {}_{-n-(\Delta(\rho)-\Delta(\sigma))}
\;\; z^{n+\Delta(\rho) -\Delta(\sigma)-\Delta(\lambda)}.
\eeq
In the above $\Delta(\lambda)$  etc. are conformal dimensions of
representation $\lambda$, i.e. $\De(\la) = \frac{C_\la}{2(N+1)}$,
$C_\la$ being the second Casimir in the $\la$-representation.
The single spinon of $\overline{N}$ representation $\phi^\alpha (z)$
on $|0\rangle$ gives the $\overline{N}$
state and the action of $H_2$ is as follows:
\bea
&& H_{2} \; \ph^\al_{-\De-n}|0\rangle=
\left[ \frac{N C_3(\overline{N})}{6(N+1)(N+2)}+N n(n-1) +(2N-1)n \right]
\ph^\al_{-\De-n}|0\rangle.
\eea
The lowest one is when $n=0$ and for $N=3$ we have the energy eigenvalue
of $\frac{1}{9}$.
Note that there is a contribution from the third Casimir of $SU(N)$
in the energy.
For the $N$-representation, the energy of a single spinon state
is obtained by changing the sign of the contribution of the third Casimir,
i.e. $C_3(N)=-C_3(\overline{N})$.
We can move up to higher energy states by considering multi-spinon states.
First let us consider the two spinon states.
The decomposition of product of two spinon states of representation
$\overline{N}$ can be written as
$\overline{N} \times \overline{N} =\frac{N(N-1)}{2} +
\overline{\frac{N(N+1)}{2}}$, i.e. we have a decomposition into
an antisymmetric one and a symmetric one.
Now we apply the action of $H_2$ on the
antisymmetric part of two spinon states to find its eigenvalues:
\bea
\Ph_{\al_{3} \cdots \al_{N}} (n_{2},n_{1}) \equiv
\ep_{\al_{1} \al_{2} \cdots \al_{N}}\;
\phi_{-\Delta_2-n_{2}}^{\al_{2}} \;\phi_{-\De_1-n_{1}}^{\al_{1}}
| 0 \rangle,  \;\;\; n_2 \geq n_1 \geq 0.
\eea
It is now straightforward to compute the action of $H_2$ on these states.
We just write down the results.
\bea
&& H_{2} \; \Ph_{\al_{3} \cdots \al_{N}} (n_{2},n_{1})=
\left[ 2\times \frac{N C_3(\overline{N})}{6(N+1)(N+2)} \right.\nonu \\
&&\hskip 1.2cm +N n_{1}(n_{1}-1)
+N(n_{2}+\De_2-\De_1)(n_{2}+\De_2-\De_1-1) \nonu\\
&&\hskip 1.2cm \left.+(2N-1)(n_{1}+n_{2}+\De_2-\De_1) \right]
\Ph_{\al_{3} \cdots \al_{N}} (n_{2},n_{1}) \nonu \\
&&\hskip 1.2cm + 2 (t^a_{1}) (t^a_{2}) \sum_{l > 0} l \;
\Ph_{ \al_{3} \cdots \al_{N}} (n_{2}+l,n_{1}-l).
\eea
In the above $t^a_i$ acts on $\ph^{\al_i}$.
Since the action of the Hamiltonian is lower triangular, we can immediately
read off the energy eigenvalues of $H_2$ given as follows:
\bea
&&E(n_1, n_2;N) = \frac{(N-1)(N-2)}{3N}+(2N-1)(n_{1}+n_{2}+\De_2-\De_1)
\nonu\\
&&\hskip 1.2cm +N\left[ n_{1}(n_{1}-1)+
(n_{2}+\De_2-\De_1)(n_{2}+\De_2-\De_1-1) \right].
\label{eq:eigenvalue}
\eea
For the $N\times N$ case, we have to change the sign of the first term.
We can apply $H_2$ on various multi-spinon states. The explicit
form of the energy eigenvalues requires explicit forms of
conformal dimensions and fusion rules of the spinon fields.

Let us illustrate this result for the simplest case, i.e $SU(3)$.
In this case, there exist three primary motifs, 1 for vacuum,
$\overline{3}, 3$ for single spinon states.
The eigenvalues for these states are $0,
\frac{1}{9}, -\frac{1}{9}$ respectively by putting $n=0$ in Eq.(25).
The excited states for $\overline{3}, 3$
can be obtained by putting $n=1$. Then the eigenvalues lead to
$\frac{46}{9}, \frac{44}{9}$ respectively.
For the two spinon state of $\overline{3} \times \overline{3}$, we have the
eigenvalues of $H_2$, $-\frac{1}{9} $ with $n_1=n_2=0$ and $\frac{26}{9}$
with $n_1=0, n_2=1$ in Eq.(28).
Similarly, the eigenvalues of $H_2$ in the case of
$ 3 \times 3 $ are $-\frac{5}{9}$ with $n_1=n_2=0$ and $\frac{22}{9}$
with $n_1=0, n_2=1$.
Other interesting two spinon cases are $\overline{N}\times N$ :
\beq
\phi^\al_{-n_2-\De_2}\; \phi_{\be,-n_1-\De_1} |0 \rangle,
\eeq
where we have $\De_2= -\frac{N-1}{2N}$ and $\De_1= \frac{N-1}{2N}$.
For $n_1=0$ and $n_2=2$ we get energy eigenvalue $2N+2$ and for
$n_1=0$ and $n_2=3$ we get the energy eigenvalue $6N+3$, consistent
with the result from the motif picture in Ref.\cite{Sch}.

We can also consider three spinon state
$\overline{\frac{N(N-1)}{2}} \times \overline{N}\times\overline{N}$
with
\beq
\phi^A\CVO{\overline{\frac{N(N-1)}{2}}}{1}{\frac{N(N-1)}{2}}
{}_{-n_3 -\De_3}
\phi^\beta \CVO{\overline{N}}{\frac{N(N-1)}{2}}{\overline{N}}
{}_{-n_2 -\De_2}
\phi^\gamma\CVO{\overline{N}}{\overline{N}}{1}
{}_{-n_1 -\De_1}|0\rangle,
\eeq
where $A=1,\cdots, \frac{N(N-1)}{2}$, and
\bea
&&\De_1 = \De(\overline{N}) =\frac{N-1}{2N},\;
\De_2 = \De\left(\frac{N(N-1)}{2}\right)
-\De(\overline{N}) =\frac{N-3}{2N}, \nonu \\
&&\De_3 = -\De\left(\frac{N(N-1)}{2}\right) =-\frac{N-2}{N}.
\eea
We can calculate the energy eigenvalue of the state explicitly for
the $N=3$ case as follows:
\bea
&&E(n_1, n_2, n_3;3) =
3 \times\frac{N C_3 (\overline{N})}{6(N+1)(N+2)}\nonu \\
&&\hskip 1.5cm +N\left[ n_{1}(n_{1}-1)+
(n_{2}+\De_2-\De_1)(n_{2}+\De_2-\De_1-1) \right. \nonu \\
&&\hskip 1.5cm +\left.(n_{3}+\De_3-\De_0)(n_{3}+\De_3-\De_0-1)\right]
\nonu \\
&&\hskip 1.5cm+ (2N-1)\left[n_{1}+n_{2}+n_3+\De_2+\De_3-\De_1-\De_0\right],
\eea
where $\De_0 = \De_1$,
and $C_3(R)$ are the third Casimirs of $R$ representation.
So for the low lying case of $n_1 =0,\ n_2 = 1,$ and $n_3 = 2$,
we get $E(0,1,2;3)=11$, consistent with the result from motif picture
\cite{Sch}.

The main observation in this letter is that it is possible to
reinterprete the Yangian generators $Q_0^a, Q_1^a$ and
the Hamiltonian $H_2$ for $SU(N)_1$ WZW CFT by exploiting the
integrability of CS model.
We can apply the forms of the Hamiltonian $H_2$ on spinon states,
to obtain some of the energy spectra, which is consistent with
the results obtained from the motif picture, justifying it.
It is also straightforward
to apply the technique
we made here to a higher Hamiltonian, for example, $H_3$ and find
what is it for WZW model from the viewpoint of CS model.
Also, the spinon basis of $SU(N)_1$ WZW model should be useful for
writing down the corresponding character formulas.
\vskip 0.5cm
\leftline{\bf Acknowledgement}

We would like to thank  H. Awata, D. Bernard, A. Jevicki, Y. Matsuo,
S. Odake, and Y. Yamada for useful discussions.  We thank K. Schoutens
for intensive discussions especially on spinons.
This work is supported in part by Ministry of Education (BSRI-95-2442),
Kyung Hee University Research Fund, Korea-Japan exchange program, and by
CTP/SNU(SRC program of KOSEF).

\end{document}